\documentclass[aip,numerical,preprint]{revtex4-1}
\usepackage{graphicx}
\usepackage{bm}
\usepackage{amsmath,amssymb}

\begin{document}

\title{Path integral Monte Carlo with importance sampling for excitons interacting with an arbitrary phonon bath}

\author{Sangwoo Shim}
\author{Al\'an Aspuru-Guzik}
\affiliation{Department of Chemistry and Chemical Biology, Harvard University, Cambridge, Massachusetts 02138, USA}

\begin{abstract}
  The reduced density matrix of excitons coupled to a phonon bath
  at a finite temperature is studied using the path integral Monte Carlo method.
  Appropriate choices of estimators and importance sampling schemes are
  crucial to the performance of the Monte Carlo simulation. We show that
  by choosing the population-normalized estimator for the reduced density
  matrix, an efficient and physically-meaningful sampling function can be
  obtained. In addition, the nonadiabatic phonon probability density is
  obtained as a byproduct during the sampling procedure.
  For importance sampling, we adopted the Metropolis-adjusted Langevin
  algorithm. The analytic expression for the gradient of the target
  probability density function associated with the population-normalized
  estimator cannot be obtained in closed form without a matrix power series.
  An approximated gradient that can be efficiently
  calculated is explored to achieve better computational scaling and efficiency.
  Application to a simple one-dimensional model system from the previous
  literature confirms the correctness of the method developed in this
  manuscript. The displaced harmonic model system within the single exciton
  manifold shows the numerically exact temperature dependence of the
  coherence and population of the excitonic system.
  The sampling scheme can be applied to an arbitrary
  anharmonic environment, such as multichromophoric systems embedded in the
  protein complex. The result of this study is expected to stimulate
  further development of real time propagation methods that satisfy
  the detailed balance condition for exciton populations.
\end{abstract}

\maketitle

\section{Introduction}
Recent 2D non-linear spectroscopy experiments suggested the existence of
long-lived quantum coherence during the electronic energy transfer process
within the Fenna-Matthews-Olson complex of green sulfur bacteria, marine algae
and plants even under
physiological conditions~\cite{Engel2007,Panitchayangkoon2010,Schlau-Cohen2012,Wong2012,Collini2010,Kolli2012}.
These results attracted a large amount of attention from
theoretical physicists and chemists. The energy transfer process
usually has been modeled as the dynamics of excitons coupled to a phonon bath in
thermal equilibrium within the single exciton manifold. This approximation leads
to the famous spin-Boson Hamiltonian. The solution of this type of 
Hamiltonian has been studied extensively. For example, by assuming a
certain relative magnitude between the reorganization energy
and coupling terms, one can obtain quantum master equations valid in
specific regimes\cite{Forster1948,May2004,Redfield1957}.
Another approximation, the Haken-Strobl-Reineker model works
in both the coherent and incoherent regimes, but incorrectly converges to the
high temperature limit in the long time even at the low temperature
~\cite{Haken1972,Haken1973}.
More recently, numerically exact approaches which interpolate both limits
have been investigated and applied to many systems of interest. Two of the most
popular methods are the hierarchical equation of motion
~\cite{Ishizaki2005,Ishizaki2009a,Ishizaki2009}
and the quasiadiabatic path integral method~\cite{Topaler1992,Makri1992}.
These methods are being actively developed, improved, and applied
to many systems of interests~\cite{Zhu2011}.

Although having been successful in many applications, many of the models
described above have assumed
the phonon bath to be a set of independent harmonic oscillators and encode
all the complexity of the bath environment in the spectral
density, which is essentially a frequency dependent distribution
of exciton-phonon coupling. However, for studying the anharmonic
effects of a very sophisticated bath environment, like the protein complexes
of photosynthesis, being
able to directly include the atomistic details of the bath structure into the
exciton dynamics has a distinct advantage.
In other words, approaches that can evaluate the influence functional first
suggested by Feynman and Vernon~\cite{Feynman1963} have more straightforward
descriptions and are applicable to arbitrary systems.
Evaluation of the exact influence functional for arbitrary environment requires 
the simulation of the full quantum dynamics, which is still not practical with currently available
computational resources. 
There have been several attempts to incorporate atomistic details of the large scale
bath by combining the exciton dynamics and molecular dynamics simulations
~\cite{Shim2012,Olbrich2010,Olbrich2011}. However, these theories are still
in their early stages and the propagation scheme used does not satisfy some fundamental properties, like the detailed balance condition at finite temperature.
In pursuit of more accurate theory, it is crucial to know the correct
asymptotic behavior in the limit of infinite time. In this context,
we decided to explore the numerically exact reduced density matrix in a
finite temperature using path integral Monte 
Carlo~\cite{Thirumalai1983,Allinger1986,Behrman1985,Cao1993} method.
Recently, Moix \textit{et al} applied path integral Monte Carlo
for the equilibrium reduced density matrix of the FMO complex within the framework
of open quantum systems~\cite{Moix2012}.

\section{Theory}
\subsection{Path integral formulation of the reduced thermal density matrix}
We want to evaluate the reduced density matrix of an excitonic system
coupled to phonons on arbitrary Born-Oppenheimer surfaces at a finite
temperature. For photosynthetic energy transfer, we usually restrict the excitons
to be within the single exciton manifold because at normal light intensity,
in average, one photon is present at a given time in the complexes of interest.
However, the formulation itself
is not limited to the single exciton manifold. The Hamiltonian operator
for such a system can be written as
\begin{align}
  \hat{H} &= \underbrace{\sum_m \int d\bm{R} \ 
        \left[ V_{m}(\bm{R}) - V_g(\bm{R}) \right]
        |m\rangle \langle m|\otimes
        |\bm{R} \rangle \langle \bm{R}|
    + \sum_{m \neq n} \int d\bm{R} \ 
        J_{mn}(\bm{R}) |m\rangle \langle n| \otimes
        |\bm{R} \rangle \langle \bm{R}|}_{\hat{H}_{\rm exc}
        = \hat{H}_S + \hat{H}_{SB}} \nonumber\\
    &+ \underbrace{|\bm{1}\rangle \langle \bm{1}| \otimes
      \left[ \hat{T} + \int d\bm{R}\ 
      V_g(\bm{R}) |\bm{R} \rangle \langle \bm{R}|
      \right]}_{\hat{H}_{B}}.
\end{align}
The Hamiltonian was written in terms of the diabatic basis
$|m,\bm{R}\rangle \equiv |m\rangle \otimes |\bm{R}\rangle$, where
$m$ is the index for the exciton state and $R$ is the phonon coordinate.
$V_g(\bm{R})$ is the potential energy surface (PES) of the phonons
in the electronic ground state and $V_m(\bm{R})$ is the PES of the phonons
in the $m$th exciton state. $\hat{T}$ is the kinetic operator of the phonons
defined as $\hat{T} = -\frac{\hbar^2}{2} \mathcal{M}^{-1} \nabla^2,$
where $\mathcal{M}$ is the mass tensor of the phonons.
This expression is generally applicable to any molecular system with
multiple potential energy surfaces.
The reduced thermal density matrix $\rho_S$ is defined as the partial trace
of the full thermal density matrix with respect to the bath degrees of freedom:
\begin{align}\label{eq:reduced_density_matrix}
  \rho_S &=  \frac{1}{Z(\beta)} {\rm Tr_B} \exp\left(-\beta \hat{H}\right)\nonumber\\
  &= \frac{1}{Z(\beta)} \int d\bm{R}_0\ \langle \bm{R}_0|\exp\left(-\beta \hat{H}\right)|\bm{R}_0\rangle,
\end{align}
where $Z(\beta)$ is the partition function of the total system. We proceed by
relying on the following identity:
\begin{align}\label{eq:discrete_decompose}
  \langle \bm{R}_0| \exp(-\beta \hat{H}) |\bm{R}_0\rangle
  &= \langle \bm{R}_0|
  \left\{ \exp\left(-\frac{\beta \hat{H}}{M}\right) \right\}^M |\bm{R}_0 \rangle\nonumber\\
  &= \int d\bm{R}_1 \int d\bm{R}_2 \cdots \int d\bm{R}_{M-1}\nonumber\\
  &\ \times
  \langle \bm{R}_0| \exp\left( -\frac{\beta \hat{H}}{M} \right) |\bm{R}_{M-1} \rangle 
  \langle \bm{R}_{M-1}| \exp\left( -\frac{\beta \hat{H}}{M} \right) |\bm{R}_{M-2} \rangle \cdots
  \nonumber\\
  &\ \times \langle \bm{R}_2| \exp\left( -\frac{\beta \hat{H}}{M} \right) |\bm{R}_1 \rangle
  \langle \bm{R}_1| \exp\left( -\frac{\beta \hat{H}}{M} \right) |\bm{R}_0 \rangle.
\end{align}

For any positive integer $M$, the expression above is exact. 
When the Trotter decomposition is applied, an imaginary timestep
$\tau \equiv \frac{\beta \hbar}{M}$ is usually defined for convenience.
Then, the thermal density matrix can be interpreted as an imaginary
time evolution. 
In the limit of an infinitesimal imaginary
timestep,
the Trotter decomposition converges to the exact result,
\begin{align}\label{eq:trotter}
  \langle \bm{R}_1| \exp\left( -\frac{\beta \hat{H}}{M} \right)
  |\bm{R}_0 \rangle
  &= \langle \bm{R}_1| \exp\left( -\tau \hat{H} / \hbar \right)
      |\bm{R}_0 \rangle \nonumber\\
  &=  \langle \bm{R}_1| e^{-\tau \hat{H}_{\rm exc} / 2\hbar }
      e^{-\tau \hat{H}_B / \hbar}
      e^{-\tau \hat{H}_{\rm exc} / 2\hbar}
    |\bm{R}_0 \rangle + O(\tau^3)\nonumber\\
  &= \int d\bm{R}_2 \int d\bm{R}_3\ 
   \langle \bm{R}_1| e^{-\tau \hat{H}_{\rm exc} / 2\hbar }
   | \bm{R}_3 \rangle \nonumber\\
   &\ \times \langle \bm{R}_3|
   e^{-\tau \hat{H}_B / \hbar}
   | \bm{R}_2 \rangle \langle \bm{R}_2|
  e^{-\tau \hat{H}_{\rm exc} / 2\hbar}
   |\bm{R}_0 \rangle + O(\tau^3).
\end{align}

Subsequently, we will recast the system part of $\hat{H}_{\rm exc}$ as a single matrix to
simplify the notation,
\begin{align}
  \hat{H}_{\rm exc} &= \sum_{m,n}\int d\bm{R}\ E_{m n}(\bm{R}) |m\rangle\langle n|
  \otimes |\bm{R}\rangle \langle \bm{R}|,\nonumber\\
  E_{m m}(\bm{R}) &= \left\{\begin{array} {rcr}
  V_m(\bm{R}) - V_g(\bm{R}) & for & m = n,\\
  J_{m n}(\bm{R}) & for & m \neq n.
  \end{array}\right.
\end{align}
With the single exciton manifold assumption, $E_{mm}$ corresponds to the
optical gap of the $m$-th site.
Now, the three terms in the integrand of the Eq.~\ref{eq:trotter} can be
written without Dirac notation,
\begin{align}\label{eq:dedirac}
  \langle \bm{R}_1| e^{-\tau \hat{H}_{\rm exc}/2\hbar} |\bm{R}_3 \rangle
  &= \delta(\bm{R}_1 - \bm{R}_3) e^{-\tau E(\bm{R}_3)/2\hbar},\nonumber\\
  \langle \bm{R}_3| e^{-\tau \hat{H}_B/\hbar} |\bm{R}_2\rangle
  &= (4\pi \tau |\lambda|)^{-1/2}
  e^{-\tau V_g(\bm{R}_3) / 2\hbar}
  e^{-(\bm{R}_3 -
  \bm{R}_2)^T\lambda^{-1}(\bm{R}_3 - \bm{R}_2)/4 \tau}
  e^{-\tau V_g(\bm{R}_2) / 2\hbar} + O(\tau^3)
  ,\nonumber\\
 \langle \bm{R}_2| e^{-\tau \hat{H}_{\rm exc}/2\hbar} |\bm{R}_0 \rangle
  &= \delta(\bm{R}_2 - \bm{R}_0) e^{-\tau E(\bm{R}_0)/2\hbar},
\end{align}
where $\lambda \equiv \frac{\hbar \mathcal{M}^{-1}}{2}$.
By the Eq.~\ref{eq:trotter} and Eq.~\ref{eq:dedirac},
\begin{align}\label{eq:buildingblock}
  \langle \bm{R}_1|\exp\left(-\frac{\beta \hat{H}}{M}\right)|\bm{R}_0\rangle
  &= (4 \pi \tau |\lambda|)^{-1/2}
    e^{-\tau V_g(\bm{R}_1) / 2\hbar}
    e^{-(\bm{R}_1 - \bm{R}_0)^T \lambda^{-1}
    (\bm{R}_1 - \bm{R}_0)/{4\tau}}
    e^{-\tau V_g(\bm{R}_0) / 2\hbar}\nonumber\\
    &\ \times e^{-\tau E(\bm{R}_1)/2\hbar} 
    e^{-\tau E(\bm{R}_0)/2\hbar} + O(\tau^3).
\end{align}
Note that Eq.~\ref{eq:buildingblock} is a matrix with the same dimension as the~
reduced density matrix of the system.
Substituting Eq.~\ref{eq:buildingblock} to Eq.~\ref{eq:reduced_density_matrix}, we obtain
\begin{align}\label{eq:mc_ground}
  \rho_S &= \frac{1}{Z(\beta)} \int d\bm{R}_0 \int d\bm{R}_1 \cdots \int d\bm{R}_{M-1}
    \nonumber\\
  &\ \times e^{-\tau E(\bm{R}_0)/2\hbar} e^{-\tau E(\bm{R}_{M-1})/\hbar} \cdots
    e^{-\tau E(\bm{R}_1)/\hbar} e^{-\tau E(\bm{R}_0)/2\hbar}
    \nonumber\\
    &\times e^{-\tau V_g(\bm{R}_0) / \hbar}
    e^{-\tau V_g(\bm{R}_1) / \hbar}
    \cdots
    e^{-\tau V_g(\bm{R}_{M-1}) / \hbar}
    \nonumber\\
    &\times e^{-(\bm{R}_0 - \bm{R}_{M-1})^T \lambda^{-1}
    (\bm{R}_0 - \bm{R}_{M-1})/ 4\tau}
    e^{-(\bm{R}_{M-1} - \bm{R}_{M-2})^T \lambda^{-1}
    (\bm{R}_{M-1} - \bm{R}_{M-2}) / 4\tau}
    \nonumber\\
    &\ \times \cdots
    \times
    e^{-(\bm{R}_1 - \bm{R}_0)^T \lambda^{-1}
    (\bm{R}_1 - \bm{R}_0)/ 4\tau}
    \nonumber\\
  &= \int d\bm{R}_0 \int d\bm{R}_1 \cdots \int d\bm{R}_{M-1}
    \nonumber\\
  &\ \times \underbrace{\frac{K}{Z(\beta)}
    e^{-\tau E(\bm{R}_0)/2\hbar} e^{-\tau E(\bm{R}_{M-1})/\hbar} \cdots
    e^{-\tau E(\bm{R}_1)/\hbar} e^{-\tau E(\bm{R}_0)/2\hbar}}_{
      \rho_{\rm PIMC}(\bm{R}_0,\cdots,\bm{R}_{M-1})}
  \nonumber\\
  &\ \times \underbrace{\frac{1}{K}
    e^{-\beta V_{\rm PIMC}(\bm{R}_0, \bm{R}_1, \cdots,
    \bm{R}_{M-1}) }}_{f_g(\bm{R}_0, \cdots, \bm{R}_{M-1})},
\end{align}
where,
\begin{align}\label{eq:v_pimc}
  V_{\rm PIMC}(\bm{R}_0, \bm{R}_1, \cdots, \bm{R}_{M-1})
  &=
  \frac{1}{M} \sum_{i=0}^{M-1} V_g(\bm{R}_i)\nonumber\\
  &\ +
  \sum_{i=0}^{M-1}\frac{M}{2\beta^2 \hbar^2}
  \{ \bm{R}_i - \bm{R}_{ {\rm mod}(i+1,M)} \}^T
  \mathcal{M}
  \{\bm{R}_i - \bm{R}_{ {\rm mod}(i+1,M)}\}.
\end{align}
The expressions above show that the reduced thermal density matrix $\rho_S$ can be evaluated as an expectation value of
$\rho_{\rm PIMC}(\bm{R}_0,\cdots,\bm{R}_{M-1})$ where the joint
probability density function of the $M$ $N$-dimensional random variables
$(\bm{R}_0, \cdots, \bm{R}_{M-1})$ is $f_g$. This type of
multidimensional integral can be efficiently evaluated using Monte
Carlo integration.
Because $f_g(\bm{R}_0, \cdots, \bm{R}_{M-1})$ is invariant to
cyclic permutation of the phonon coordinate, usually the averaged estimator
$\rho_{\rm PIMC}$ over the cyclic permutation is used in the actual Monte
Carlo evaluation:
\begin{align}
  \rho_{\rm \overline{PIMC}} = \frac{1}{M} \sum_{i=0}^{M-1}
  \rho_{\rm PIMC}(\bm{R}_i, \bm{R}_{ {\rm mod}(i+1,M)},
  \cdots,\bm{R}_{ {\rm mod}(i+M-1,M)}).
\end{align}

\subsection{Population-normalized estimator and importance sampling}
In the previous approach described in Eq.~\ref{eq:mc_ground}, the phonon
coordinates are sampled according the electronic ground state PES.
The estimator should converge to the target quantity in the long time limit,
taking into account the discretization error. As long as
$f_g(\bm{R}_0,\cdots,\bm{R}_{M-1})$ is positive definite everywhere in the
phonon space, the sampling efficiency depends on the selection of the
probability density. Obviously, the actual distribution of the phonon
coordinate depends heavily on the excited state PES.
Therefore, the Monte Carlo points coordinates sampled according to
the reduced dynamics of the bath by taking the partial trace with respect to
the \emph{exciton degrees of freedom}, as explored in multiple surface path
integral Monte Carlo approaches, are expected to give the better
estimates. This choice of the probability density reweights the
estimator in the following way:
\begin{align}
  f_I(\bm{R}_0, \cdots, \bm{R}_{M-1}) &= 
  {\rm Tr_S} \left[ \rho_{\rm \overline{PIMC}} (\bm{R}_0, \cdots, \bm{R}_{M-1}) \right]
  f_g(\bm{R}_0, \cdots, \bm{R}_{M-1}),\nonumber\\
  \rho_{I}(\bm{R}_0, \cdots, \bm{R}_{M-1}) &= 
  \frac{\rho_{\rm \overline{PIMC}} (\bm{R}_0, \cdots, \bm{R}_{M-1})}{{\rm Tr_S} \left[ \rho_{\rm \overline{PIMC}} (\bm{R}_0, \cdots, \bm{R}_{M-1}) \right]}.
\end{align}
In the expression above, we call $\rho_I(\bm{R}_0, \cdots, \bm{R}_{M-1})$ the population normalized
estimator for the reduced density matrix because the sum of its populations
is always constrained to be 1. The effective energy gap term of
$-\frac{1}{\beta} \log {\rm Tr} \rho_{\rm \overline{PIMC}}(\bm{R}_0,\cdots,\bm{R}_{M-1})$ was added to the
Eq.~\ref{eq:v_pimc} to enable the phonons
follow the excited state dynamics depending on the exciton state $\rho_S.$
For the estimator of the reduced density matrix in Eq.~\ref{eq:mc_ground},
the normalization must obtained by the estimates of its diagonal elements,
leading to more uncertainties in the coherence. However, the population-normalized
estimator preserves the correct normalization by construction, and
does not suffer from any additional uncertainty.

Local gradient information can improve the efficiency and scaling of the
sampling procedure by means of a gradient-based approach such as the
Metropolis-adjusted Langevin algorithm (MALA).
\cite{Robert2004,Pillai2011}
However, the exact closed form of the gradient of the effective energy gap term,
$\log {\rm Tr_S} \rho_{\rm \overline{PIMC}} (\bm{R}_0, \cdots, \bm{R}_{M-1})$
can only be expressed as a function of a power series of matrices.
Nevertheless, with the following approximation:
\begin{align}
\sum_{k=0}^n A^k B A^{n-k} \approx \sum_{k=0}^n \frac{1}{2^n} {\binom{n}{k}} A^k B A^{n-k},
\end{align}
an accurate approximated of the gradient can be obtained and employed in the sampling procedure,
\begin{align}
  \frac{\partial}{\partial R_{ij}} \log {\rm Tr_S} \left[ \rho_{\rm \overline{PIMC}} (\bm{R}_0, \cdots, \bm{R}_{M-1}) \right]
  &= \frac{ {\rm Tr_S} \left[ \frac{\partial}{\partial R_{ij}} \rho_{\rm \overline{PIMC}} (\bm{R}_0, \cdots, \bm{R}_{M-1}) \right] }{ {\rm Tr_S} \left[ \rho_{\rm \overline{PIMC}} (\bm{R}_0, \cdots, \bm{R}_{M-1}) \right]}\nonumber\\
  &\approx \frac{ {\rm Tr_S} \left[ -\frac{\tau}{2\hbar}
  \frac{\partial E(\bm{R}_i)}{\partial {R}_{ij}}
  \rho_{\rm \overline{PIMC}} (\bm{R}_0, \cdots, \bm{R}_{M-1}) \right] }{ {\rm Tr_S} \left[ \rho_{\rm \overline{PIMC}} (\bm{R}_0, \cdots, \bm{R}_{M-1}) \right]},
  \nonumber\\
  \nabla_i \log f_g(\bm{R}_0, \cdots, \bm{R}_{M-1})
  &= -\frac{\beta}{M} \nabla_i V_g(\bm{R}_i)
  \nonumber\\
  &\ + \frac{M}{2\beta \hbar^2} \mathcal{M} ( \bm{R}_{ {\rm mod}(i+1,M)} +
  \bm{R}_{ {\rm mod}(i-1,M)} - 2 \bm{R}_i),\nonumber\\
  \mu_i(\bm{R}_0, \cdots, \bm{R}_{M-1})
  &= \frac{ {\rm Tr_S} \left[ -\frac{\tau}{2\hbar}
  \frac{\partial E(\bm{R}_i)}{\partial {R}_{ij}}
  \rho_{\rm \overline{PIMC}} (\bm{R}_0, \cdots, \bm{R}_{M-1}) \right] }{ {\rm Tr_S} \left[ \rho_{\rm \overline{PIMC}} (\bm{R}_0, \cdots, \bm{R}_{M-1}) \right]}
  \nonumber\\
  &+ \nabla_i \log f_g(\bm{R}_0, \cdots, \bm{R}_{M-1})\nonumber\\
  &\approx \nabla_i \log f_I(\bm{R}_0, \cdots, \bm{R}_{M-1}).
\end{align}
Here, $\nabla_i$ is the  gradient operator with respect to $\bm{R}_i$.

Note that if we choose an appropriate Metropolis criterion, no bias in the
distribution is introduced even with the approximate gradient~\cite{AspuruGuzik2003}.
Firstly, a trial move $\bm{R}'_i$ obtained by
\begin{align}
  \bm{R}'_i = \bm{R}_i +\mu_i ( \bm{R}_0, \cdots, \bm{R}_{M-1}) \Delta t +
  \xi_i \sqrt{\Delta t},
\end{align}
where $\Delta t$ is the timestep for the Monte Carlo step and $\xi_i$ is a $N$-dimensional vector of independent
standard Gaussian random variables. Then, $R_i'$ is probabilistically accepted 
according to the acceptance ratio,
\begin{align}
  \frac{f_I(\bm{R}'_0, \cdots, \bm{R}'_{M-1})}{f_I(\bm{R}_0, \cdots, \bm{R}_{M-1})} \times
  \frac{\prod_{i=0}^{M-1}\exp \left[ - \frac{|\bm{R}'_i - \{ \bm{R}_i + \mu_i(\bm{R}_0, \cdots, \bm{R}_{M-1}) \} |^2}{2\Delta t} \right]}
         {\prod_{i=0}^{M-1}\exp \left[ - \frac{|\bm{R}_i - \{ \bm{R}'_i + \mu_i(\bm{R}'_0, \cdots, \bm{R}'_{M-1}) \} |^2}{2\Delta t} \right]}.
\end{align}
The Monte Carlo timestep $\Delta t$ is only a tunable parameter for the
Monte Carlo sampling procedure and not related to the physics of the
simulated system.

\section{Application}
\subsection{Alexander's 1D test model}
Our formulation is equivalent to the multiple electronic state extension of
matrix multiplication path integral (MMPI) method of Alexander~\cite{Thirumalai1983,Alexander2001}
when the population normalized estimator is chosen and
only the vibrational degrees of freedom are considered. Therefore, the
1D model employed in Ref.~\citenum{Alexander2001} was calculated to test the validity of our method.
The elements of the electronic Hamiltonian in this model are given by,
\begin{align}
  V_{11}(x) &= \frac{1}{2} k_{11}(x-x_{11}) ^ 2 + \varepsilon_{11},
  \nonumber\\
  V_{22}(x) &= \frac{1}{2} k_{22}(x-x_{22}) ^ 2 + \varepsilon_{22},
  \nonumber\\
  V_{12}(x) &= c \exp \left[ -\alpha (x - x_{12}) ^2 \right],
\end{align}

\begin{figure}
  \includegraphics[width=0.5\textwidth]{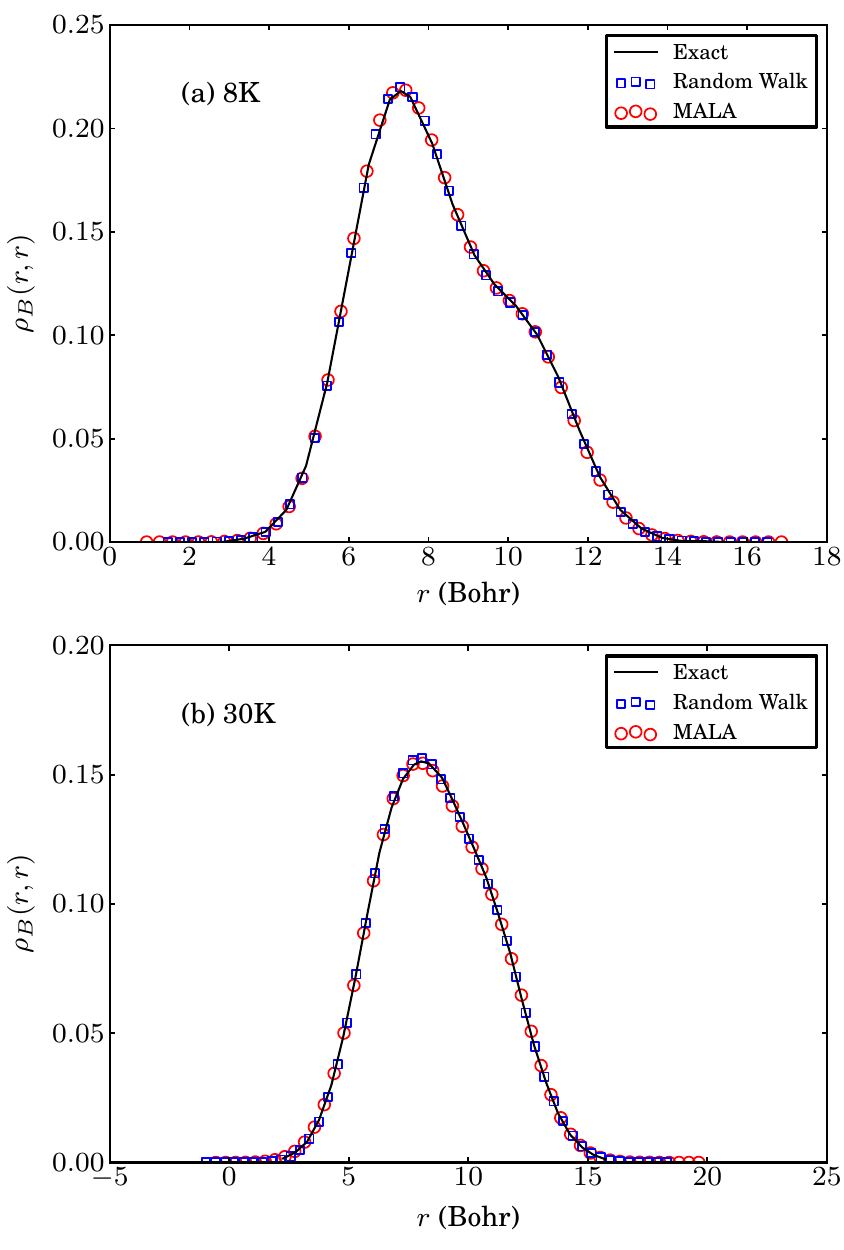}
  \caption{The estimated nuclear probability densities of Alexander's model~\cite{Alexander2001}
   at (a) 8K and (b) 30K. For path integral Monte Carlo simulations,
   densities were obtained by histograms with 50 bins. The discretization
   number of 8 was enough to converge to the exact probability densitiies.}
  \label{fig:alexanderPhononDensity}
\end{figure}

\begin{table}
  \begin{tabular}{c c}
    \hline
    \hline
    Parameters & Value \\
    \hline
    \hline
    $k_{11}$ & $4 \times 10^{-5}$ \\
    \hline
    $k_{22}$ & $3.2 \times 10^{-5}$ \\
    \hline
    $x_{11}$ & $7$ \\
    \hline
    $x_{22}$ & $10.5$ \\
    \hline
    $\varepsilon_{11}$ & $0$ \\
    \hline
    $\varepsilon_{22}$ & $2.2782 \times 10^{-5}$ \\
    \hline
    $c$ & $5 \times 10^{-5}$ \\
    \hline
    $\alpha$ & $0.4$ \\
    \hline
    $x_{12}$ & $8.75$ \\
    \hline
    $m$ & $3.6743 \times 10^3$\\
    \hline
  \end{tabular}
  \caption{Summary of the parameters for the model system by Alexander \textit{et al}~\cite{Alexander2001}.
    All values are given in atomic units.}
  \label{tbl:alexander_params}
\end{table}
The total nuclear probability density evaluated as histograms from
the Metropolis random walk and MALA simulations are compared to the grid-based
result from Alexander \textit{et al.}~\cite{Alexander2001} in Fig.
\ref{fig:alexanderPhononDensity}. The distributions converged to the exact
probability density after $2 \times 10^7$ steps with 8 beads at both
temperatures of 8K and 30K.

\subsection{Model of a chromophore heterodimer with displaced harmonic oscillators}\label{sec:dimer}
To test the proposed method, a system of two chromophores in a photosynthetic complex was modeled using
displaced harmonic oscillator model. In this model, the ground and excited
electronic states of the monomer are modeled as harmonic oscillators with
different displacement, but the same harmonic constant~\cite{May2004}.
The thermal reduced density matrix 
was calculated within the single exciton manifold.
The Hamiltonian for this model is then given as follows:
\begin{align}
  V_g(x_1, x_2) &= \frac{1}{2} (k_1 x_1 ^2 + k_2 x_2 ^2),
  \nonumber\\
  V_e(x_1, x_2) &=  \left( \begin{array}{cc}
    \frac{1}{2} k_1 \{ (x_1 - d_1) ^ 2 - x_1^2 \} + \varepsilon_1 & J \\
    J & \frac{1}{2} k_2 \{ (x_2 - d_2) ^ 2 - x_2^2 + \varepsilon_2\} 
  \end{array} \right),
  \nonumber\\
  \mathcal{M} &= \left( \begin{array}{cc}
    m_1 & 0 \\
    0 & m_2 \end{array} \right).
\end{align}
\begin{table}
  \begin{tabular}{c  c}
  \hline
  \hline
  Parameter & Value \\
  \hline
  \hline
  $k_1$ & $2.227817 \times 10^{-3}$ \\
  \hline
  $k_2$ & $2.227817 \times 10^{-3}$ \\
  \hline
  $d_1$ & $3.00000$ \\
  \hline
  $d_2$ & $2.00000$ \\
  \hline
  $\varepsilon_1$ & $8.064745 \times 10^{-2}$ \\
  \hline
  $\varepsilon_2$ & $7.976238 \times 10^{-2}$ \\
  \hline
  $J$ & $ -4.738588 \times 10^{-4}$ \\
  \hline
  $m_1$ & $3.418218 \times 10^6$ \\
  \hline
  $m_2$ & $3.418218 \times 10^6$ \\
  \hline
\end{tabular}
\caption{Summary of the parameters for the displaced harmonic oscillator model
used in Sec.~\ref{sec:dimer}.
    All values are given in atomic units.}
\label{tbl:dimer_params}
\end{table}
Some of the parameters were set according to our molecular dynamics/quantum
chemistry calculation of the FMO complex~\cite{Shim2012}. The parameter values are
listed in table \ref{tbl:dimer_params}.

The model system was simulated at seven different temperatures ranging from 30K to 300K
with a number of beads (discretization number) of 4, 8, 16, 32 and 64.
The number of timesteps propagated in each simulation was $4 \times 10^7$.
The value of each timestep was tuned so that the acceptance ratio of the
MALA run is close to 0.574, and 0.234 for the Metropolis random walk as
maintaining these acceptance ratio is known to provide most efficient
sampling~\cite{Pillai2011}. We used non-overlapping batch means~\cite{Flegal2010}
with a batch size of $10^6$ to estimate the standard error of the correlated samples.
The batch size was adjusted so that the null hypothesis of uncorrelated batches
was not rejected by using Ljung-Box test~\cite{Ljung1978} at a significance level of $5\%$.

\begin{figure}
  \includegraphics[width=0.5\textwidth]{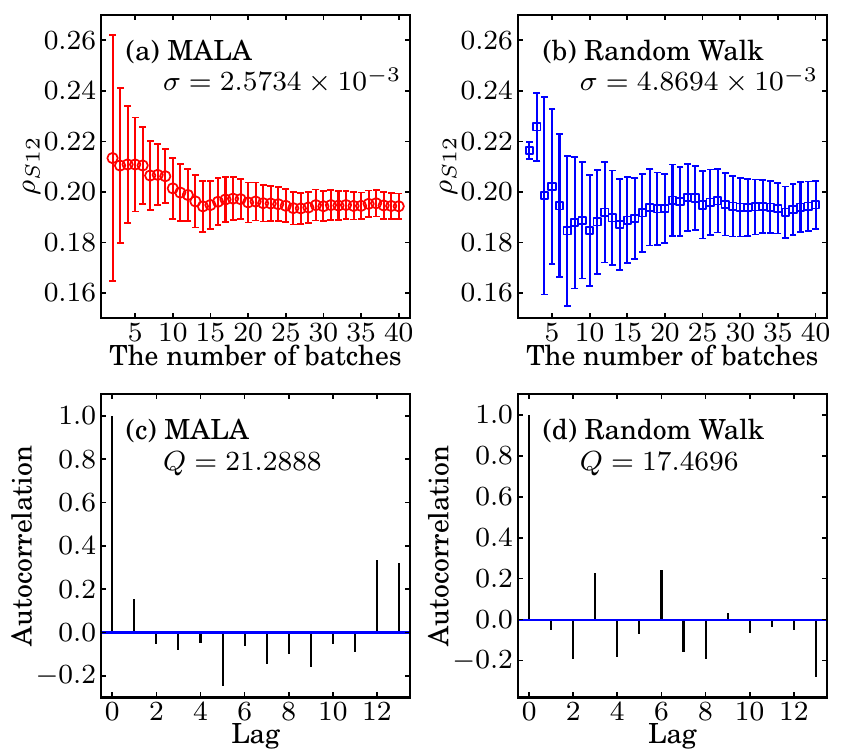}
  \caption{Estimates of (1,2) matrix elements of the thermal reduced density
   matrix evaluated using MALA and Metropolis random walk at 77K with 64 beads.
  MALA estimate has a smaller confidence interval thus a more accurate estimate
  than that of the Metropolis random walk. The error bar indicates the
  95\% confidence interval evaluated with the batch means.
  The $0.95$ quantile of the $\chi^2$ distribution with 13 degrees of freedom
  is 22.362 and both Ljung-Box statistics ($Q$) are smaller. Thus, the uncorrelation hypothesis
  is not rejected in both cases at the 5\% significance level.}
  \label{fig:dimerBatchError}
\end{figure}

\begin{figure}
  \includegraphics[width=0.5\textwidth]{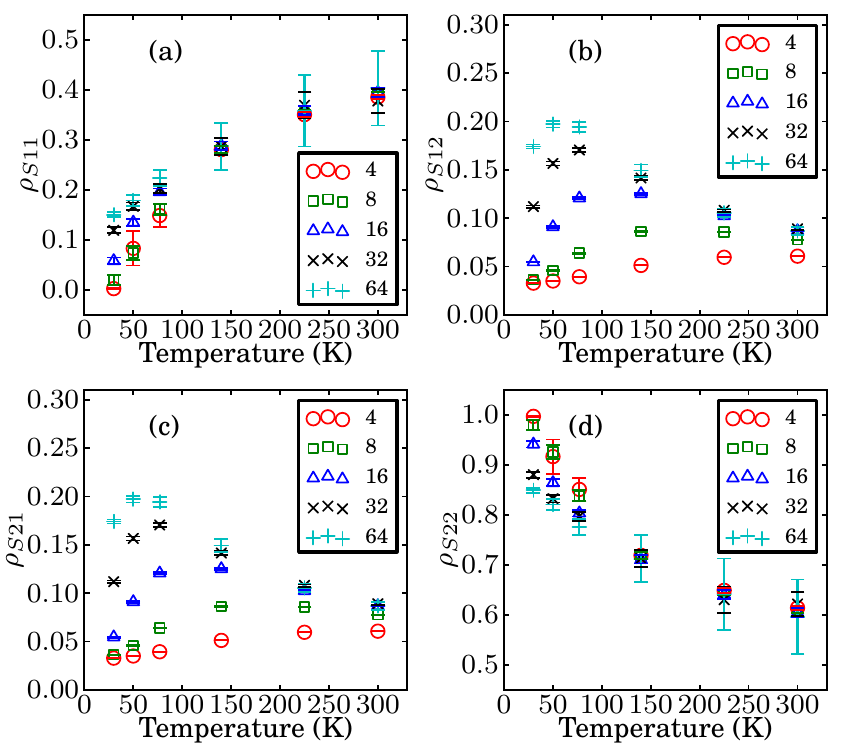}
  \caption{Estimates of matrix elements of the thermal reduced density
   matrix evaluated at 30K, 50K, 77K, 140K, 225K and 300K with different
  discretization numbers of 4, 8 and 16 using MALA.
  (a) is the (1,1) element,
  (b), (c) and (d) are (1,2), (2,1) and (2,2) elements, respectively. The error bar indicates
   the 95\% confidence interval evaluated with the batch means.}
  \label{fig:dimerTempDepend}
\end{figure}

\begin{figure}
  \includegraphics[width=0.5\textwidth]{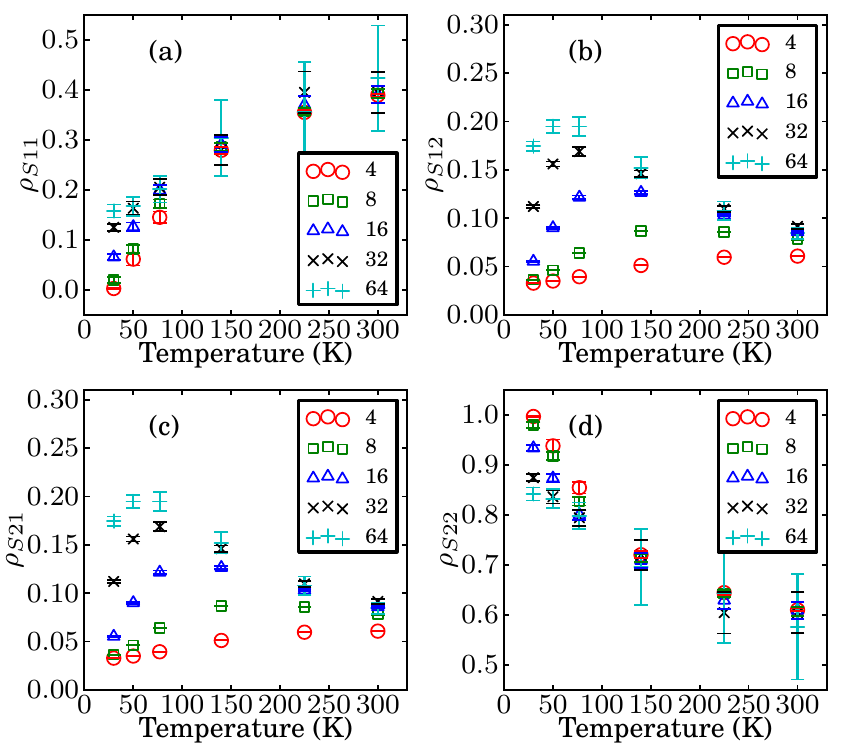}
  \caption{Estimates of matrix elements of the thermal reduced density
   matrix evaluated at 30K, 50K, 77K, 140K, 225K and 300K with different
  discretization numbers of 4, 8 and 16 using random walk Metropolis.
  (a) is the (1,1) element,
  (b), (c) and (d) are (1,2), (2,1) and (2,2) elements, respectively. The error bar indicates
   the 95\% confidence interval evaluated with the batch means.}
  \label{fig:dimerTempDependRandomWalk}
\end{figure}

\begin{figure}
  \includegraphics[width=\textwidth]{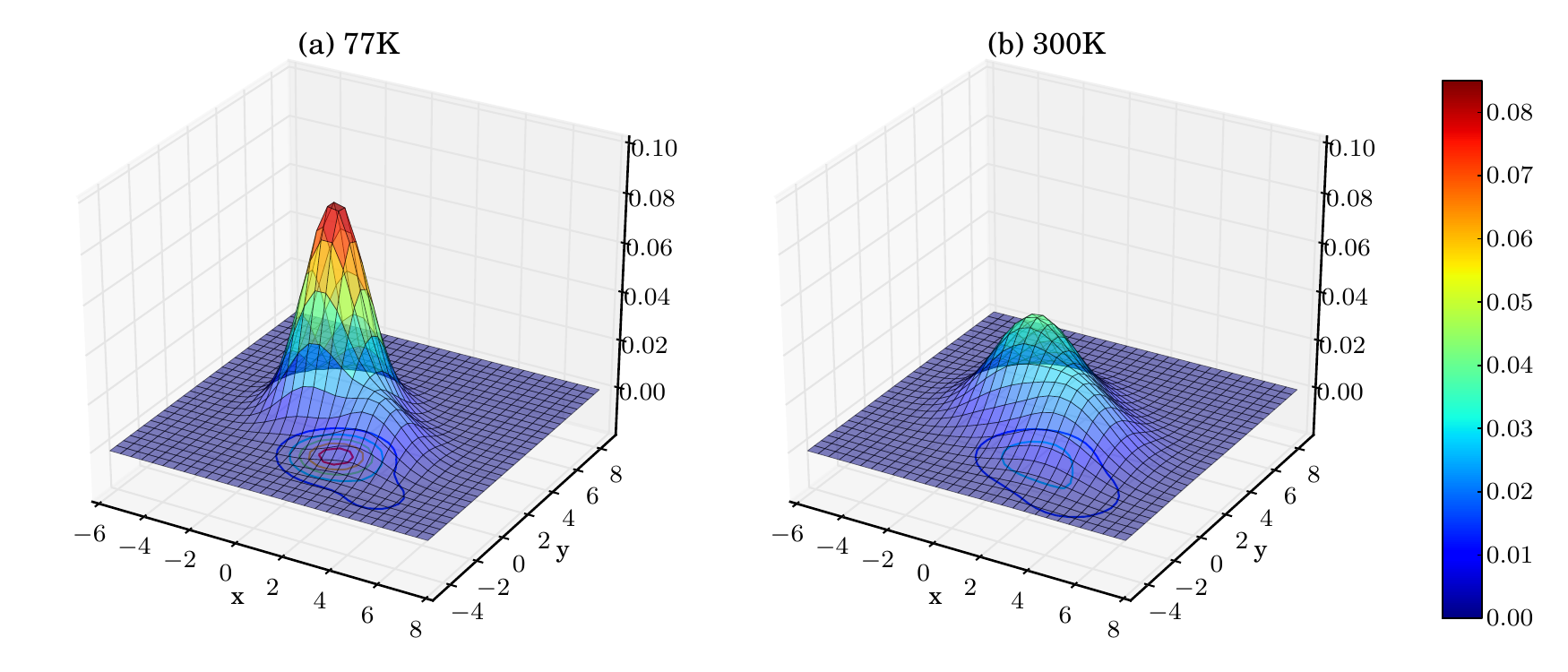}
  \caption{The phonon probability density function evaluated at (a) 77K and (b) 300K
  with 16 beads using MALA. At the lower temperature, the contribution of the
  exciton with lower energy at $(0, 2)$ becomes larger. Therefore, the population 
  differenece becomes more distinct, as can be seen in the temperature dependence
  of the exciton population in Fig. \ref{fig:dimerTempDepend}.}
  \label{fig:dimerPhononDensity}
\end{figure}

As shown in Fig.~\ref{fig:dimerBatchError}, the standard error of the simulation
decreases modestly as the number of Monte Carlo steps increases.
Fig. \ref{fig:dimerTempDepend} shows the temperature dependence of the
estimates of reduced density matrix elements as a function of various discretization
numbers using MALA.
Although the Metropolis random walk simulation gives a smaller confidence interval
for the 4 bead case, MALA provides better estimates as the dimension of the
sample space increases. The Metropolis random walk result is given in
Fig.~\ref{fig:dimerTempDependRandomWalk}.
While the population of the low energy site decreases
as the temperature increases, the quantum coherence does not
monotonically decrease.
We believe that this pheonomenon is an artifact of an insufficient
discretization number at low temperatures. As can be seen in Fig.
\ref{fig:dimerTempDepend}, 64 or more beads are needed for the coherence
to converge at 77K, while 16 beads are enough at 300K with acceptable
accuracy. This is a well known limition of imaginary time path integral
Monte Carlo simulations.
Figure \ref{fig:dimerPhononDensity} shows the probability density function of the
phonon coordinate at 77K and 300K. The population difference in the reduced
density matrix is reflected to the difference in the probability mass of the
two diabatic potential energy minimum at $(3, 0)$ and $(0, 2).$

\section{Conclusion}
We explore a method for obtaining the thermal reduced density
matrix of an exciton system coupled to an arbitrary phonon bath for
path integral Monte Carlo simulation.
Note that our scheme is closely related to the path integral Monte Carlo
simulation for nonadiabatic systems for vibrational coherence
\cite{Schwieters1999,Alexander2001,Schmidt2007}.
Although the phonon state can be obtained as a byproduct, we mainly focused on the
evaluation of the reduced density matrix of the excitonic system
to explore the asymptotic behavior of the populations and coherences in this paper.
In addition, we implemented an importance sampling scheme
for better spatial scaling and sampling efficiency. 
Although the path integral Monte Carlo cannot evaluate the real time
evolution of density matrices, the method gives the exact asymptotic values
with all quantum effects from both the system and bath environments if a
sufficient number of beads are used. We believe that in some of the cases
where the bath has a nontrivial coupling to the system, or the non-Markovianity
of the bath manifests very strongly, treating the environment around the
system of interest as a set of harmonic oscillators is not sufficient.
If this is the case, the system should be studied in its entirety.
We are trying to develop a real time propagation scheme to treat the system
exactly, and the bath semiclassically. The method studied in this paper
offers a foundation for it by providing the correct asymptotic behaviors.

\begin{acknowledgments}
S.S. thanks the Samsung Scholarship for financial support. 
This work was supported by the Defense Advanced Research
Project Agency Award No. N66001-10-4060 and by the
Center of Excitonics, and Energy Frontier Research Center
funded by the U.S. Department of Energy,
Office of Science, and Office of Basic Energy Sciences
under Award No. DE-SC0001088.
A.A.-G. also acknowledges generous support from 
the Alfred P. Sloan and the Camille and Henry Dreyfus foundations.
\end{acknowledgments}

\bibliographystyle{jcp}
\bibliography{pimc-oqs-ng}
\end{document}